\documentclass[a4paper,11pt]{article}

\usepackage[labelfont=bf,labelsep=period]{caption}
\usepackage[nodayofweek]{datetime}
\usepackage{float}
\usepackage{graphicx}
\usepackage{adjustbox}
\usepackage{tikz}
\usepackage{hyperref}
\usepackage[utf8]{inputenc}
\usepackage{numbertabbing}
\usepackage{url}
\usepackage{xcolor}
\usepackage{xspace}
\usepackage{algpseudocode}
\usepackage{subcaption}
\usepackage[margin=2.5cm]{geometry}

\usepackage{tikz}
\usetikzlibrary{automata}
\usetikzlibrary{decorations.pathreplacing,angles,quotes, decorations.pathmorphing} 
\usetikzlibrary{math}

\definecolor{Colour1}{HTML}{ffffd4}
\definecolor{Colour2}{HTML}{fee391}
\definecolor{Colour3}{HTML}{fec44f}
\definecolor{Colour4}{HTML}{fe9929}
\definecolor{Colour5}{HTML}{d95f0e}
\definecolor{Colour6}{HTML}{993404}

\usetikzlibrary{external}
\usepackage{amsmath} 
\allowdisplaybreaks[2]          
\usepackage{amssymb} 
\usepackage{amsthm} 
\usepackage{dsfont}
\newtheoremstyle{plain-boldhead}
  {\topsep}
  {\topsep}
  {\itshape}
  {}
  {\bfseries}
  {.}
  { }
  {\thmname{#1}\thmnumber{ #2}\thmnote{ (\bfseries #3)}}
\newtheoremstyle{definition-boldhead}
  {\topsep}
  {\topsep}
  {\normalfont}
  {}
  {\bfseries}
  {.}
  { }
  {\thmname{#1}\thmnumber{ #2}\thmnote{ (\bfseries #3)}}
\theoremstyle{plain-boldhead}
\newtheorem{theorem}{Theorem}

\newtheorem{lemma}[theorem]{Lemma}

\theoremstyle{definition-boldhead}
\newtheorem{definition}{Definition}

\newdateformat{simple}{\THEDAY\ \monthname[\THEMONTH]\ \THEYEAR}
\simple

\floatstyle{ruled}
\newfloat{algo}{htbp}{algo}
\floatname{algo}{Algorithm}

\newfloat{module}{htbp}{module}
\floatname{module}{Module}

\def \ifempty#1{\def\temp{#1} \ifx\temp\empty }

\newcommand{\str}[1]{\textsc{#1}}
\newcommand{\var}[1]{\textit{#1}}
\newcommand{\op}[1]{\textsl{#1}}
\newcommand{\msg}[2]{\ensuremath{\ifempty{#2} [\str{#1}] \else [\str{#1}, {#2}] \fi}}

\newcommand{\CA}{\ensuremath{\mathcal{A}}\xspace}

\newcommand{\CD}{\ensuremath{\mathcal{D}}\xspace}

\newcommand{\CH}{\ensuremath{\mathcal{H}}\xspace}

\newcommand{\CN}{\ensuremath{\mathcal{N}}\xspace}

\newcommand{\CR}{\ensuremath{\mathcal{R}}\xspace}

\newcommand{\upon}{\textbf{upon}\xspace}

\title{DAG it off: Latency Prefers No Common Coins}
\author{Ignacio {Amores-Sesar} \\
Aarhus University\\
\url{amores-sesar@cs.au.dk} 
\and 
Viktor Grøndal \\
Aarhus University\\
\url{au727671@uni.au.dk} 
\and
Adam Holmgård \\
Aarhus University\\
\url{au717856@uni.au.dk} 
\and 
Mads Ottendal \\
Aarhus University\\
\url{au729914@uni.au.dk} 
}

\begin{document}

\maketitle

\begin{abstract}
We introduce Black Marlin, the first Directed Acyclic Graph (DAG)-based Byzantine atomic broadcast protocol in a partially synchronous setting that successfully forgoes the reliable broadcast and common coin primitives while delivering transactions every round. Black Marlin achieves the optimal latency of 3 rounds of communication (4.25 with Byzantine faults)  while maintaining optimal communication and amortized communication complexities. We present a formal security analysis of the protocol, accompanied by empirical evidence that Black Marlin outperforms state-of-the-art DAG-based protocols in both throughput and latency.
\end{abstract}

\section{Introduction}
\label{sec:intro}
Scalability poses a significant challenge for consensus protocols. Paradoxically, increasing the number of participants in consensus degrades its performance rather than enhancing it.

Traditional consensus protocols, such as Paxos~\cite{DBLP:journals/tocs/Lamport98}, PBFT~\cite{DBLP:conf/osdi/CastroL99}, and HotStuff~\cite{DBLP:conf/podc/YinMRGA19}, primarily adopt a leader-based approach. In this model, a designated leader proposes a value, which the remaining participants validate. This design results in an uneven distribution of workload, with the leader becoming a performance bottleneck. If we add more parties to a consensus protocol, we simply increase the communication load of the leader, degrading the performance of the protocol. Furthermore, if the leader behaves in a Byzantine manner, the system must invoke a view-change mechanism to replace it. These view-changes are sensitive to timing assumptions and become progressively more expensive~\cite{DBLP:books/daglib/0025983} as the number of parties increases.

To address these limitations, Keidar et al.~\cite{DBLP:conf/podc/KeidarKNS21} introduced DAG-Rider, an elegant asynchronous solution that leverages the common core abstraction~\cite{DBLP:conf/stoc/CanettiR93}. This approach enables every participant to act as a leader in the protocol. These instances are interwoven into a directed acyclic graph (DAG), on top of which a total ordering is achieved using a common coin~\cite{DBLP:books/daglib/0025983}, applied at the end of every \emph{wave}: a set of four consecutive rounds.

DAG-Rider demonstrates remarkable throughput compared to leader-based protocols, but this comes at the cost of significantly increased latency. The primary cause of this latency is the repeated use of reliable broadcast~\cite{DBLP:books/daglib/0025983} across each of the four rounds required to implement the common core primitive~\cite{DBLP:conf/stoc/CanettiR93}. This triggered the effort of the research community~\cite{DBLP:conf/eurosys/DanezisKSS22,DBLP:conf/ccs/SpiegelmanGSK22,DBLP:conf/wdag/KeidarNPS23,DBLP:conf/sosp/GiridharanSAAC24,DBLP:journals/iacr/ShresthaKN24,DBLP:conf/ndss/BabelCDKKKST25,DBLP:conf/fc/SpiegelmanAGL24,DBLP:conf/nsdi/Arun0S0S25} aimed at mitigating this latency overhead. These contributions generally fall into two broad categories.

On the one hand, Cordial Miners~\cite{DBLP:conf/wdag/KeidarNPS23} replaces the reliable broadcast rounds with bare-message communications, reducing communication overhead at the cost of increasing the wave length to five rounds (three in partial synchrony). On the other hand, a majority of the works~\cite{DBLP:conf/eurosys/DanezisKSS22,DBLP:conf/ccs/SpiegelmanGSK22,DBLP:conf/sosp/GiridharanSAAC24,DBLP:journals/iacr/ShresthaKN24,DBLP:conf/ndss/BabelCDKKKST25,DBLP:conf/fc/SpiegelmanAGL24,DBLP:conf/nsdi/Arun0S0S25} pursue orthogonal improvements, either through more efficient use of the common coin or by shortening the length of the waves under partial synchronous assumptions. These approaches have collectively succeeded in halving the latency of DAG-based protocols, marking a significant advance in the state of the art. However, their reliance on the reliable broadcast primitive continues to be a limiting factor. Moreover, to the best of our knowledge, no existing protocol has yet successfully elected an anchor in every round.

These are precisely the two limitations addressed in this paper. We introduce Black Marlin, the first partially synchronous, DAG-based protocol that elects an anchor in every round without relying on reliable broadcast. The simplest—yet most impactful—innovation is the substitution of the common coin with a deterministic round-robin mechanism. This substitution is valid because the FLP impossibility~\cite{DBLP:journals/jacm/FischerLP85} only applies in asynchronous settings.

The primary advantage of the round-robin approach is that it allows block creation to be conditioned on the reception of the anchor block from a given round. To handle scenarios in which a malicious anchor fails to behave correctly, Black Marlin employs timeouts, thereby preventing deadlocks. This conditional block creation enables Black Marlin to elect an anchor in every round, fulfilling the long-standing goal of the research on DAG-based protocols.

Moreover, by operating in a partially synchronous model, Black Marlin is able to adapt ideas from Cordial Miners~\cite{DBLP:conf/wdag/KeidarNPS23} to eliminate the need for reliable broadcast, using only digital signatures without the need for extra rounds of communication. The trade-off for achieving anchor election without reliable broadcast is a small constant delay and a minor reduction in commitment probability under Byzantine behavior, while maintaining optimal message complexity in both cases.

Specifically, Black Marlin commits the anchor of round $r-2$ upon completing round $r$ with probability $4/9$ under adversarial conditions. This probability increases to 1 when all participants behave honestly. As a result, the expected latency is 4.25 communication rounds under Byzantine behavior and the optimal 3 rounds in the honest case.

Additionally, timeouts are only triggered during asynchronous periods or when facing Byzantine faults. Thus, Black Marlin remains a reactive protocol, effectively leveraging the advantages of both synchronous and asynchronous models. We prove that Black Marlin implements atomic broadcast~\cite{DBLP:books/daglib/0025983} together with all the performance results mentioned above. Furthermore, we benchmark Black Marlin against Bullshark, showing its practicality.

\subsection{Structure}
The paper is structured as follows. Section~\ref{sec:intro} briefly describes the paper. Section~\ref{sec:related} details the state of the art of DAG-based protocols. Section~\ref{sec:model} sets the notation and key concepts used in the paper. Section~\ref{sec:protocol} describes Black Marlin in full detail. Section~\ref{sec:analysis} shows that Black Marlin implements atomic broadcast and studies its complexities. Section~\ref{sec:bench} describes a comparison of the throughput and latency of Black Marlin and Bullshark.

\section{Related work}
\label{sec:related}

DAG-based protocols have constituted a significant transformation in the field of consensus over the past few years.

This revolution began in the permissionless setting, with protocols such as Avalanche~\cite{DBLP:journals/corr/abs-1906-08936} and IOTA~\cite{DBLP:journals/candie/PopovSF19}. However, these protocols were initially introduced without thorough security analyses. As a result, several vulnerabilities were later discovered~\cite{DBLP:journals/iotj/CullenFKS20,DBLP:conf/opodis/Amores-SesarCT22,DBLP:conf/sirocco/AmoresSesarCS24}, which limited their success despite recent work demonstrating their optimality~\cite{DBLP:conf/esorics/Amores-SesarC24}.

DAG-based consensus protocols were later extended to the permissioned setting, most notably with DAG-Rider by Keidar et al.~\cite{DBLP:conf/podc/KeidarKNS21}. DAG-Rider was the first protocol to leverage the properties of the common core primitive~\cite{DBLP:conf/stoc/CanettiR93} to implement atomic broadcast in asynchronous networks. It outperforms traditional leader-based protocols~\cite{DBLP:journals/tocs/Lamport98,DBLP:conf/osdi/CastroL99,DBLP:conf/podc/YinMRGA19} in terms of throughput. However, DAG-Rider requires on average six instances of reliable broadcast~\cite{DBLP:journals/jacm/BrachaT85}, resulting in 18 rounds of communication. Since its introduction, several works have aimed to address the latency limitations of DAG-Rider.

On one hand, Cordial Miners~\cite{DBLP:conf/wdag/KeidarNPS23} reduces latency to an average of 7.5 communication rounds by replacing reliable broadcast with bare-message communication. The tradeoff of this approach is that its new communication paradigm requires an additional round per wave to compensate for the properties that reliable broadcast provides. When a party creates a block in Cordial Miners, it must include any blocks the recipient may have missed. We adopt this idea in the design of Black Marlin.

On the other hand, a widely adopted strategy to improve DAG-Rider’s latency has involved either assuming synchronous networks or modifying the core and coin primitives. Narwhal and Tusk~\cite{DBLP:conf/eurosys/DanezisKSS22} introduced a clever transaction preprocessing mechanism, where transactions are grouped into batches, so only the hashes of these batches are submitted to consensus. This batching is delegated to specialized \emph{workers}, allowing the system to scale dynamically with transaction load. However, Narwhal and Tusk lack a rigorous security analysis, and liveness vulnerabilities have been identified~\cite{DBLP:journals/iacr/Shoup24a}.

Bullshark~\cite{DBLP:conf/ccs/SpiegelmanGSK22}, which builds on top of Narwhal and Tusk, also introduces a variant for partial synchrony that improves latency to just three instances of reliable broadcast, about nine rounds of communication, on average. Bullshark is particularly relevant to our work, as its authors provide a public implementation~\cite{bullsharkgit}\footnote{Bullshark has been chosen over more recent frameworks such as Mysticeti~\cite{mysticetigit} due to its closer similarity to Black Marlin and the timing of this project.}, enabling a fair, side-by-side comparison with Black Marlin. It is worth noting that both Narwhal and Tusk, as well as Bullshark, rely on reliable broadcast as a fundamental building block.

More recently, several protocols have been proposed targeting latency. These works largely follow the above trajectory. Mysticeti~\cite{DBLP:conf/ndss/BabelCDKKKST25} reduces latency further by employing consistent broadcast, achieving an average-case latency of only three instances~\cite{DBLP:books/daglib/0025983}. Sailfish~\cite{DBLP:journals/iacr/ShresthaKN24} also improves latency, though it continues to rely on reliable broadcast. Similarly, Shoal and Shoal++~\cite{DBLP:conf/fc/SpiegelmanAGL24} provide enhancements in anchor selection, but still depend on reliable broadcast.

To the best of our knowledge, Black Marlin is the first DAG-based protocol to achieve optimal latency in partial synchrony, while electing an anchor in every round, a crucial feature for minimizing the latency of non-anchor blocks without relying on reliable broadcast. We also highlight that Black Marlin achieves these properties while preserving simplicity.

\section{Model}
\label{sec:model}
\subsection{Notation} We consider a set $\CN=\{0,...,n-1\}$ of $n$ parties running a round-based algorithm. In each round, these parties create blocks that form the vertices of a directed acyclic graph $(\var{DAG},\CR)$ with references that constitute the edges. Edge $(B_1,B_2)\in \CR$ if $B_1$ includes a reference to $B_2$. A block $B= [r,i,\var{refs},\var{refs}']$ is a tuple formed by a round number $r$, an identifier for party $i$, a set of strong references $\var{refs}$, and a set of weak references $\var{refs}'$. For simplicity, we omit the set of transactions and the signature from the block, as these transactions play no role in the consensus mechanism. Given a block $B$, $\op{creator}(B)$ denotes the party that created $B$, $\op{round}(B)$ denotes the round in which $B$ was created. The variable $\var{DAG}(r)$ denotes the set of block $B'\in\var{DAG}$ such that $\op{round}(B')=r$. A party selected by a round-robin mechanism is referred to as \emph{anchor}, blocks produced by the anchor party are referred to as \emph{anchor} blocks.

\subsection{Adversary and Network}
We model parties as \emph{interactive Turing machines} (\emph{ITM}). An interactive Turing machine is a Turing machine with an input and an output tape that allows the Turing machine to communicate with other Turing machines and make decisions based on the content of their input tape. The adversary is modeled as another ITM that corrupts up to $f$ parties at the beginning of the execution, the actual number of corruptions is denoted by $t\leq f< \frac{n}{3}$. These corrupted parties obey the adversary; i.e., they may diverge from the execution of the protocol. Corrupted parties are often referred to as \emph{Byzantine} and non-corrupted as \emph{honest}.

We assume a partial synchronous communication model~\cite{DBLP:journals/jacm/DworkLS88} on point-to-point links. The network begins in an asynchronous state and turns synchronous after a global stabilization time (GST). After this point, every message sent is delivered with a delay of up to $\Delta$.
The adversary not only decides the global stabilization time but also when the message is made available to the party within the $\Delta$ time limit. We also assume that the local computations are instantaneous, this assumption can easily be relaxed by factoring the computation of the honest parties in the network delay. We assume the existence of a public key infrastructure. We also assume that every block and a message produced by every party is signed, and that the signature scheme is secure.

\subsection{Atomic broadcast} We model our protocol as \emph{atomic broadcast}. Our atomic broadcast primitive is accessed with the events $\op{ab-broadcast}(B,i,r)$ and outputs events $\op{ab-deliver}(B,j,r)$. The event $\op{ab-broadcast}(B,i,r)$ can only be triggered by party $i$, but $\op{ab-deliver}(B,j,r)$ may be output by every party.

\begin{definition}[Atomic broadcast]
  \label{def:abbroadcast} 
  A protocol solves \emph{atomic broadcast} if it satisfies the following properties, except with negligible probability:
      \begin{description}
      \item[Validity.] If an honest party $i$ invokes $\op{ab-broadcast}(B,i,r)$ after the global stabilization time, then party $i$ eventually outputs $\op{ab-deliver}(B,i,r)$.
      \item[Agreement.] If an honest party outputs $\op{ab-deliver}(B,i,r)$, then all honest parties eventually output $\op{ab-deliver}(B,i,r)$.
      \item[Integrity.] For any party $i$ and round $r$, every honest party outputs $\op{ab-deliver}(B,i,r)$ at most once regardless of $B$. Moreover, if $i$ is honest, then $i$ has invoked $\op{ab-broadcast}(B,i,r)$.
      \item[Total order.] If an honest party outputs $\op{ab-deliver}(B,i,r)$ before $\op{ab-deliver}(B',j,r')$, then no honest party outputs $\op{ab-deliver}(B',j,r')$ before  $\op{ab-deliver}(B,i,r)$.   
      \end{description}
  \end{definition}

Note that our definition of atomic broadcast is adapted to the partial synchrony model; the validity property applies to blocks that are broadcast after the network becomes synchronous. However, the safety properties are satisfied during both synchrony and asynchrony.

\section{Protocol}
\label{sec:protocol}

\begin{figure}[H]
\begin{center}
\begin{adjustbox}{max width=\textwidth}
\includegraphics[width=\linewidth]{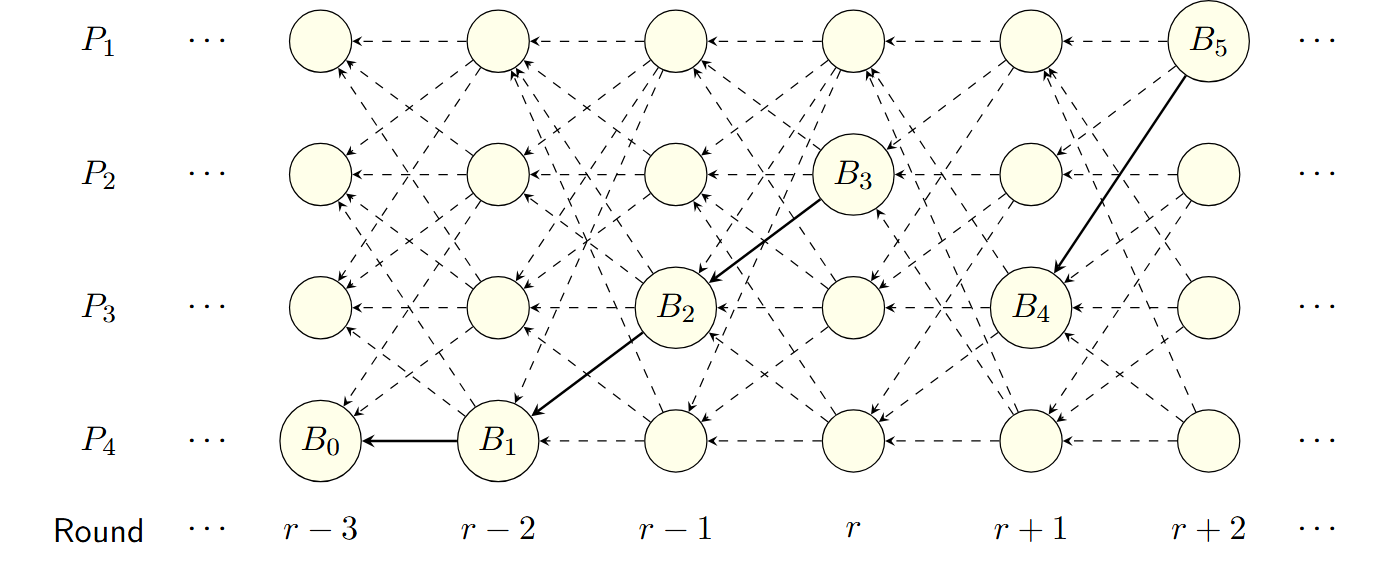}
\end{adjustbox}
\end{center}
\caption{An example of execution of Black Marlin (Algorithm~\ref{algo:main}) with $f=1$ and $n=4$.  Every block has references to at least $3$ blocks from the previous round. The blocks $B_0,...,B_5$ are the anchor blocks for the corresponding round. Note that blocks $B_0$, $B_1$, $B_2$, and $B_3$ have enough support  (i.e., at least 3 references) to be committed. However, $B_3\notin\op{strong}(B_4)$, thus $B_0$, $B_1$, and $B_2$, satisfy the commitment rule (L\ref{l:fun_commit1}–\ref{l:fun_commit2}) while $B_3$ does not. After the global stabilization time, this situation cannot occur without malicious behavior, as $B_3$ would satisfy $B_3\in\op{past}(B_4)$. This situation has been constructed to illustrate the commitment rule.}

\label{fig:links}
\end{figure}

Black Marlin (Algorithm~\ref{algo:main}) proceeds in rounds. In round~$r$, honest parties wait to receive blocks from at least $n!-!f$ parties. They then check if an anchor for round~$r$ has been received and the anchors for rounds $r-2$ and $r-1$ appear in the past of at least $n-f$ blocks from rounds $r-1$ and $r$, respectively. If both hold, a new block is created referencing all received blocks, and the party advances to round $r+1$. Otherwise, the party waits until the conditions are met or a timeout triggers block creation. These timeouts ensure liveness under Byzantine anchors and are only used when progress is blocked, allowing the protocol to be responsive. Figure~\ref{fig:links} illustrates an example execution.

\subsection{Validity predicates} A block from round $r \neq 1$ and creator $j$ is valid if it is signed by party $j$ and contains references to valid blocks created by at least $n-f$ parties from round $r-1$. Since blocks in round $r=1$ do not reference any previous blocks, a block is valid if it has been signed by its creator. A history $\mathcal{H}$ is valid if all the blocks contained in it are also valid.

 When receiving a message $\msg{block}{B', \mathcal{H}', j, r'}$ (L\ref{l:receive}), party $i$ checks the validity of both the block $B'$ and the history $\mathcal{H}'$, and adds the block $B'$ along with every unknown block from the history $\mathcal{H}'$ to its local view ($\var{DAG}$). Party $i$ also updates party $j$’s local history (L\ref{l:history_update}), used solely to optimize message complexity by sending only strictly necessary blocks.

\subsection{Initialization} The first round of the protocol differs from the rest as there are no blocks from earlier rounds. Party $i$ creates a block with an empty set of references and invokes the event $\op{ab-broadcast}(B,i)$ before sending the message $\msg{block}{B, \{\},i,0}$ to every party and starting a timeout with time $2\Delta$. Party $i$ then moves to the next round, a standard round.

\subsection{Round structure} Every round beyond the first shares the same structure. Consider an honest party $i$ in round $r$, i.e., it has just produced a block corresponding to round $r$. Party $i$ waits to receive messages $\msg{block}{B, \CH, j, r'}$ from at least $n-f$ different parties for some round $r' \geq r$. Once sufficient messages for such a round have been received, party $i$ checks whether it has received a message $\msg{block}{B', \CH', j, r'}$ with $j$ being the anchor for the current round, and verifies that the previous two anchors have enough support, or if a timeout has been triggered, to conclude the round. This dual condition to start a new round allows the protocol to be responsive, i.e., it runs at the speed of the actual network delay instead of the estimated network delay $\Delta$ without deadlocking when an anchor party behaves maliciously.

Party $i$ concludes the round by performing the following sequential operations. First, $i$ attempts the delivery of blocks (L\ref{l:attempt}); this function is described in its own paragraph below. Secondly, party $i$ creates a new block $B$ corresponding to the next round $r' + 1$, including strong references $\var{refs}$ to every block in $\var{DAG}(r')$ (L\ref{l:references1}–\ref{l:references2}). A set of weak references $\var{refs}'$ to blocks from previous rounds are also included; this secondary set is time-bounded, preventing the usual garbage collection problem associated with these solutions. Party $i$ concludes these operations by sending $B$ to every other party, together with the history of the parties (L\ref{l:send1}). Note that party $i$ may skip one or multiple rounds if an appropriate quorum is received.

\begin{figure}[H]
\begin{center}
\begin{adjustbox}{max width=\textwidth}
\includegraphics[width=\linewidth]{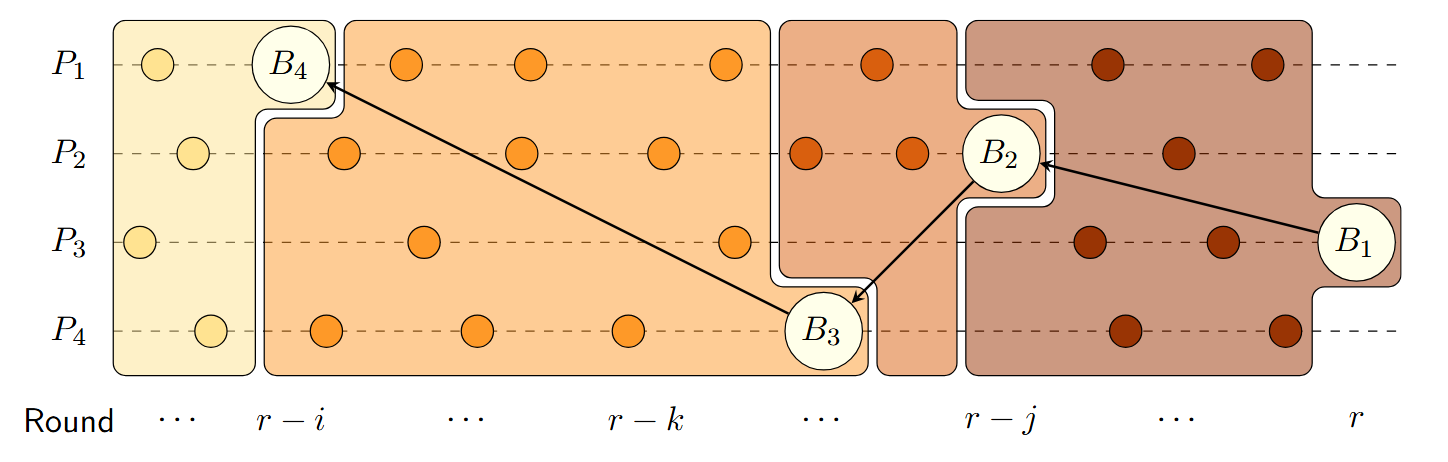}
\end{adjustbox}
\end{center}
\caption{Illustration of the commit rule of Black Marlin. When a party invokes the commit function on an anchor block $B_1$, it will recursively commit earlier uncommitted anchor blocks first. The subscript numbers on the blocks indicate the order in which the commit function is called. In this process, blocks $B_1$ $B_2$, $B_3$, and $B_4$ are $\op{ab-delivered}$, in the reverse order of their commitment. Before committing any anchor block, all non-delivered blocks in their causal history are deterministically sorted and committed. In the illustration, these correspond to the smaller vertices contained within the history boxes associated with each anchor block.}
\label{fig:BMCommitRule}
\end{figure}

\subsection{Delivery} The delivery function determines which blocks are delivered and in what order, based on the local view of the DAG. Upon concluding round $r$, party $i$ attempts the commitment of the anchor block $B$ from round $r-2$ (L\ref{l:fun_attempt1}--\ref{l:fun_attempt2}). $B$ is an anchor block if its creator has been elected by the round-robin mechanism. However, it is only committed if it is supported by at least $n-f$ parties in round $r-1$ and including an anchor of round $r-1$ $B'$, and $B'$ is also supported by at least $n-f$ parties in round $r$. These conditions prevent honest parties from committing different blocks when the anchor party is Byzantine.

If this check succeeds, party $i$ calls the function $\op{commit}(B)$ (L\ref{l:fun_commit1}–\ref{l:fun_commit2}), which iterates through the local view of the DAG. Party $i$ searches for a non-delivered block $B'$ from a higher round reachable through strong references from $B$. If such a block $B'$ exists, the function $\op{commit}(B')$ is called recursively. If no such block exists, party $i$ deterministically orders the blocks in $\op{past}(B)$ before invoking $\op{ab-deliver}$ on them. In case of conflicting blocks from the same party and round, only the first block—determined by the deterministic ordering above—is $\op{ab-delivered}$. Figure~\ref{fig:BMCommitRule} shows an example of the commitment rule.

\subsection{Functions}
Black Marlin (Algorithm~\ref{algo:main}) uses special functions as described in the paragraphs above. For the sake of the reader, we summarize the functions together in this section.

\begin{algo*}
  \vbox{
  \begin{numbertabbing}
    xxxx\=xxxx\=xxxx\=xxxx\=xxxx\=xxxx\=MMMMMMMMMMMMMMMMMMM\=\kill
    $\CN$\` set of parties\label{}\\
    $r \gets 0$ \` current round\label{}\\
    $\var{DAG}\gets \emptyset$ \` local view of the DAG\label{}\\
    $\CD\gets\emptyset$ \` set of delivered blocks\label{}\\
    $\var{refs}\gets \emptyset$\` set of strong references\label{}\\
    $\var{refs}'\gets \emptyset$\` set of weak references\label{}\\
    $\op{history}[j\in\CN]\gets \emptyset$\` set of blocks known by party $j$\label{}\\
    \\ 

    \textbf{function} $\op{quorum}(r)$\label{}\\
    \> \textbf{return} $|\op{parties}(\var{DAG}(r))|\geq n-f$\label{}\\
    \\
    \textbf{function} $\op{anchor}(r)$\label{}\\
    \> \textbf{return} $\exists B \in\var{DAG}(r): \op{creator}(B)=\op{RR}(r)$\label{}\\
    \\
    \textbf{function} $\op{suppAnchor}(r)$\label{}\\
    \> \textbf{return} $ \exists B \in\var{DAG}(r): \op{creator}(B)=\op{RR}(r) \land \op{supp}(B)\geq n-f$\label{}\\
    \\

    \textbf{function} $\op{delivery}(r)$ \label{l:fun_attempt1}\` //attempt a block in $r-2$\\
    \> $B\gets\op{RR}(r-2)$\label{}\` //if multiple consider them all\\
    \> \textbf{if} $\op{supp}(B)\geq n-f \land \exists B' \in \var{DAG}(r-1): $\label{l:anchor} \\
    \>\>$\op{creator}(B')= \op{RR}(r-1)\land B\in\op{strong}(B') \land \op{supp}(B')\geq n-f$ \textbf{then}\\
    \>\> $\op{commit}(B)$ \label{l:fun_attempt2}\\
    \\

    \textbf{function} $\op{commit}(B)$ \label{l:fun_commit1}\\
     \>$\CA\gets\op{strong}(B)\setminus \CD$\label{}\` //ignore already delivered blocks\\
     \>\textbf{if} $\CA\neq\emptyset$ \textbf{then}\label{}\\
     \>\> \textbf{if} $|\op{maxAnchor}(\CA)|=1$ \textbf{then}\label{}\\
     \>\>\> $B'\gets \op{maxAnchor}(\CA)$\label{p:reduce}\` //anchor from the highest round\\
     \>\> \textbf{else}\label{}\\
     \>\>\> $B'\gets A\in \op{maxAnchor}(\CA): |\op{round}(A)-\op{round}(\op{maxAnchor}(\op{strong}(A)))|$\label{}\\
    \>\>\>  is minimal and break ties deterministically \\ 
     \>\> $\op{commit}(B')$\label{}\\
     \> \textbf{for} $B'\in\tau(\op{past}(B)\setminus \CD)$ \textbf{do}\label{}\` // with $\tau$ any deterministic topological sorting\\
     \>\>\textbf{if} $\nexists B^*\in \CD:\op{creator}(B')= \op{creator}(B^*) \land \op{round}(B')=\op{round}(B^*)$ \textbf{then}\label{}\\
     \>\>\> $\op{ab-deliver}(B',\op{creator}(B'),\op{round}(B'))$\label{}\\
     \>\>\> $\CD\gets \CD \cup {B'}$\label{}\\
     \> $\op{ab-deliver}(B,\op{creator}(B),\op{round}(B))$\label{}\\
     \> $\CD\gets \CD \cup {B}$\label{}\\
     \> \textbf{return}\label{l:fun_commit2}
    \end{numbertabbing}
  }
  \caption{State and functions (party $i$)}
  \label{algo:functions}
\end{algo*}

Given a block $B$, the function $\op{creator}(B)$ returns the party that created and signed $B$, and the function $\op{round}(B)$ returns the round in which $B$ was created. The validity predicate $V(B)$ returns $1$ if $B$ has at least $n-f$ blocks from different parties and no two blocks from the same party corresponding to $\op{round}(B)-1$ in its references, and $B$ has been signed by its creator. When invoked with a set of blocks $\CH$, the validity predicate $V(\CH)$ returns $1$ if every block $B'\in\CH$ satisfies $V(B')$. The function $\op{past}(B)$ returns the set of blocks reachable from $B$ through strong and weak links, and the function $\op{strong}(B)$ returns the set of blocks reachable from $B$ through strong links, $B$ is not part of the output of either function. The function $\op{supp}(B)$ counts the number of parties $i$ that created a block $B_i$ in round $\op{round}(B)+1$ such that $B\in\op{strong}(B_i)$ and no other block from the same creator and round is in $\op{strong}(B_i)$. The function $H(B)$ is a collision-resistant hash function. The operation $\op{setTimeout}(r,\rho)$ starts a timeout of duration $\rho$ corresponding to round $r$, whereas $\op{timeout}(r)$ triggers when the timeout is completed and $\op{clearTimeout}(r)$ removes any active timeout corresponding to round $r$. The function $\op{time}()$ returns the local time of the party and when input with a round $r$, $\op{time}(r)$ returns the local time when the party adopted round $r$. In the case that the party never adopted round $r$, $\op{time}(r)$ returns the time when the party adopted the smallest round $r'>r$. The function $\op{RR}(r)$ takes a round as an input and returns the party by the round-robin mechanism in round $r$, a block from this party is also referred to as \emph{anchor}. In case multiple blocks exist, e.g. due to Byzantine behavior, the function returns both blocks. The variable $\var{DAG}(r)$ returns the set of blocks from round $r$ in the local view of the party. The operation $\op{maxAnchor}(\CA)$ takes a set of blocks $\CA$ as an input and returns the anchor from the highest round. Given a round $r$, the function $\op{quorum}(r)$ returns 1 if there are blocks from at least $n-f$ parties in $\var{DAG}(r)$. The function $\op{anchor}(r)$ returns 1 if there is a block created by the anchor of $r$ in $\var{DAG}(r)$, and $\op{suppAnchor}(r)$ returns 1 if there is an anchor block in $r$ that is supported by at least $n-f$ parties in $\var{DAG}(r+1)$.

\section{Analysis}
\label{sec:analysis}
Black Marlin (Algorithm~\ref{algo:main}) is designed for partially synchronous networks, which means that it may not be alive during the asynchronous period. However, its safety properties are still guaranteed during this period. This fact is reflected in the analysis below.
\subsection{Safety}
The results in this section apply to every block, even during the asynchronous phase.

\begin{lemma}
\label{lemma:honestmessage}
    Every honest message $\msg{block}{B', \CH',j,r}$ received by an honest party $i$ is valid, i.e., satisfies line L\ref{l:valid} in the view of $i$.
\end{lemma}

\begin{proof}
     A message $\msg{block}{B', \CH',j,r}$ is valid if block $B'$ and history $\CH'$ satisfy the validity predicate $V$. If the sender is honest, both $B'$ and $\CH'$ satisfy the validity predicate. Thus, it is only left to prove that party $i$ has enough information to verify the validity predicate. The construction of the history $\CH'$ guarantees this fact.
\end{proof}

Lemma~\ref{lemma:honestmessage} applies whenever an honest party sends a message. For the sake of readability, we omit explicit references to this lemma each time an honest message is sent.

\begin{lemma}
    \label{lemma:unique}
    Given two honest parties $i$ and $j$ and blocks $B_i$ and $B_j$ from round $r$, if $B_i$ satisfies line~L\ref{l:anchor} in the view of party $i$ and $B_j$ satisfies line~L\ref{l:anchor} in the view of party $j$, then $B_i=B_j$.
\end{lemma}

\begin{proof}
    Given a block $B_i$ created by party $k$ in round $r$ that satisfies line L\ref{l:anchor} in the view of some party, it follows that $\op{creator}(B_i) = \op{RR}(r)$ and $\op{supp}(B_i) \geq n - f$. Recall that, by definition of the support function, an honest party supports at most one block per party per round. Therefore, if blocks $B_i$ and $B_j$ both satisfy line L\ref{l:anchor} in the views of parties $i$ and $j$ respectively, then the following condition must be satisfied $\op{creator}(B_i) = \op{creator}(B_j) = \op{RR}(r)$ and $\op{supp}(B_i), \op{supp}(B_j) \geq n - f$.
    Given the corruption threshold condition of $n \geq 3f + 1$, it is impossible for two distinct blocks to each have support from at least $n - f$ parties. Hence, it must be the case that $B_i = B_j$.
\end{proof}
Lemma~\ref{lemma:unique} states that at most one block is honestly committed per round. Furthermore, these committed blocks create a chain in the DAG, as we prove in the following lemmas.

\begin{algo*}
  \vbox{
  \begin{numbertabbing}
    xxxx\=xxxx\=xxxx\=xxxx\=xxxx\=xxxx\=MMMMMMMMMMMMMMMMMMM\=\kill
    \upon \op{init} \textbf{do} \label{}\\
    \> $\op{ab-broadcast}(B,i,1)$\label{}\\
    \> \op{send} message \msg{block}{B, \{\},i,r} to every party \label{}\\
    \> $r\gets 1$\label{}\\
    \> $\op{setTimeout}(1,2\Delta)$\label{}\\
    \\
    \upon $\exists r'\geq r: \op{quorum}(r')$ \textbf{do} \label{}\\
    \> $\op{setTimeout}(r',2\Delta)$\label{}\\ 
    \\
    
    \upon$\exists r'\geq r: \op{quorum}(r')\ \land$  \label{l:newround}\\
    \>$(\op{anchor}(r')\land \op{suppAnchor}(r'-1) \land \op{suppAnchor}(r'-2) ) \lor \op{timeout}(r'))$ \textbf{do}\\
    \> $\op{clearTimeout}(r')$\label{}\\
    \> \op{delivery}(r')\label{l:attempt}\\
    \> $\var{refs}\gets \emptyset$ \label{l:references1}\\
    \> $r_\ell\gets \min \{r'\in\mathbb{N}: \ |\op{time}()-\op{time}(r)|\leq 3\Delta\}$\label{}\\
    \> $\var{weak}\gets \bigcup_{r^*=r_\ell}^{r'-1}\op{DAG}(r^*)$\label{}\\
    \> \textbf{for} $j\in \op{parties}(\var{DAG}(r))$ \textbf{do}\label{}\\
    \>\> $B'\gets B\in \var{DAG}(r): \op{creator}(B')=j$\` if multiple $B'$, select any\label{}\\
    \>\>$\var{refs}\gets \var{refs}\cup H(B')$\label{}\\
    \>\>$\var{weak}\gets \var{weak}\setminus\op{past}(B')$\label{}\\
    \> $\var{refs}'\gets \cup_{B^*\in\var{weak}}H(B^*)$ \label{l:references2}\\
    \> $r\gets r'+1$\label{}\\
    \> $B\gets [r,i,\var{refs},\var{refs}']$\label{}\\
    \> $\op{ab-broadcast}(B,i,r)$\label{}\\
    \> \textbf{for} $j\in \CN$ \textbf{do}\label{l:send1}\\
    \>\> \op{send} message \msg{block}{B, \var{DAG}\setminus\op{history}[j],i,r} to party $j$ \label{l:send_history}\\
    \>\> $\op{history}[j]\gets \var{DAG}$\label{}\\
    \\

    \upon \op{receiving}  message \msg{block}{B',\CH',j,r'} \textbf{do} \label{l:receive}\\
    \> \textbf{if} $V(B')\land V(\CH')$ \textbf{then} \label{l:valid}\\
    \>\> $\var{DAG}\gets \var{DAG}\cup \{B'\}\cup \CH'$\label{}\\
    \>\> $\op{history}[j]\gets \op{history}[j] \cup \CH'\cup \{B'\}$\label{l:history_update}
    \end{numbertabbing}
  }
  \caption{Black Marlin (party $i$)}
  \label{algo:main}
\end{algo*}

\begin{lemma}
    \label{lemma:every}
    Consider an honest party $i$ and the maximal round $r$ such that $\var{DAG}(r)\neq \emptyset$ in the view of party $i$, then $|\var{DAG}(r')|\geq n-f$ for every round $r'<r$.
\end{lemma}
\begin{proof}
    Consider the maximal round $r$ and a block $B\in\var{DAG}(r)$. Since $B$ is a valid block, there exists valid blocks from at least $n-f$ parties in $\op{strong}(B)$ corresponding to round $r-1$. Thus, $|\var{DAG}(r-1)|\geq n-f$. Now, consider any block in $\var{DAG}(r-1)$ and iterate recursively. We conclude that $|\var{DAG}(r')|\geq n-f$ for every round $r'<r$.
\end{proof}

In contrast to traditional leader-based protocols, a block in DAG-based protocols must be in the past of an overwhelming majority of blocks to ensure its delivery. The following result formalizes this requirement for committed blocks.

\begin{lemma}
    \label{lemma:follow1}
    If a block $B$ from round $r$ satisfies line L\ref{l:anchor} in the view of some honest party $i$, then every valid block $B'$ from round $r'\geq r+2$ satisfies $B\in\op{past}(B')$.
\end{lemma}

\begin{proof}

    By definition of line L\ref{l:anchor}, $\op{supp}(B)\geq n-f$, thus $B\in\op{past}(B')$ of blocks $B'$ from at least $n-f$ different parties in round $r+1$. We proceed by induction over $\rho:= r'-r$:
    \begin{itemize}
        \item  Let $B'$ be any block from round $r + 2$. By construction, there are blocks from at least $n-f$ in $\var{DAG}(r+1)$, and up to $f$ of them could be Byzantine. Block $B$ is in the past of blocks from round $r+1$ from at least $n-f$ different parties, thus $B\in\op{past}(B')$. We conclude $B\in\op{past}(B')$ of any block $B'$ from round $r+2$.
        
        \item Assume that every block $B_0$ from round $r+\rho_0$ satisfies $B\in\op{past}(B_0)$ and let $B'$ be a block from round $r'=r+\rho_0+1$. We have that $B\in\op{past}(B')$, since there are at least $n-f$ blocks in $\op{past}(B')$ and every block $B_0$ from round $r+\rho_0$ fulfills $B\in\op{past}(B_0)$.
    \end{itemize}

    We conclude that any block $B$ from round $r$ that satisfies line L\ref{l:anchor} in the view of some honest party, then every block $B'$ from round $r'\geq r+2$ satisfies $B\in\op{past}(B')$.
\end{proof}

\begin{lemma}
    \label{lemma:follow2}
    If honest parties $i$ and $j$ commit blocks $B_i$ and $B_j$ respectively, then $B_i=B_j$, $B_i\in\op{past}(B_j)$, or $B_j\in\op{past}(B_i)$.
\end{lemma}
\begin{proof}
    Let $\op{round}(B_i)$ and $\op{round}(B_j)$ denote the rounds in which parties $i$ and $j$ commit blocks $B_i$ and $B_j$, respectively. We analyze three cases based on the values of these rounds:
    \begin{itemize}
    \item \textbf{Case 1:} $\op{round}(B_i) = \op{round}(B_j)$.\\
    By Lemma~\ref{lemma:unique}, at most one block can be committed per round. Hence, $B_i = B_j$.

    \item \textbf{Case 2:} $\op{round}(B_i) < \op{round}(B_j)$.  We further distinguish two sub-cases:
    \begin{itemize}
        \item \textbf{Case 2a:} $\op{round}(B_j) = \op{round}(B_i) + 1$.\\
        By the second condition in line~L\ref{l:anchor}, there exists a block $B_j' \in \var{DAG}(\op{round}(B_j))$ such that $\op{creator}(B_j)=\op{creator}(B_j')$ and $B_i\in \op{strong}(B_j')$. Since Byzantine parties may create multiple blocks, we cannot immediately conclude that $B_j = B_j'$. However, the third condition of line~L\ref{l:anchor} ensures that both $B_j$ and $B_j'$ satisfy $\op{supp}(B_j), \op{supp}(B_j') \geq n - f$, which, by Lemma~\ref{lemma:unique}, implies $B_j = B_j'$. Thus, $B_i \in \op{past}(B_j)$.

        \item \textbf{Case 2b:} $\op{round}(B_i) +1 < \op{round}(B_j) $.\\
        In this case, Lemma~\ref{lemma:follow1} guarantees that $B_i \in \op{past}(B_j)$.
    \end{itemize}

    \item \textbf{Case 3:} $\op{round}(B_i) > \op{round}(B_j)$.\\
    This is symmetric to Case 2 and follows by applying the same reasoning, yielding $B_j \in \op{past}(B_i)$.
\end{itemize}
    We conclude that any two blocks committed by honest parties satisfy that both parties commit the same block or blocks in the past of each other.
\end{proof}
This chain of committed blocks is common across honest parties, thus defining a partial order common to every honest party.

\subsection{Liveness}
\label{sec:liveness}
Liveness is guaranteed only after the global stabilization time. For this reason, the majority of the results presented here apply only to blocks that are $\op{ab-broadcast}$ after time $\var{GST}$.

\begin{lemma}
    \label{lemma:bounds1}
    If an honest party invokes $\op{ab-broadcast}(B,i,r)$ at time $T\geq \var{GST}$, then every honest party $j$ that invokes $\op{ab-broadcast}(B',j,r)$, does it before time $T+3\Delta$.
\end{lemma}
\begin{proof}
    Assume that party $i$ invokes $\op{ab-broadcast}(B, i, r)$ at time $T\geq \var{GST}$. Then, according to line~L\ref{l:send_history}, $i$ sends a message of the form $\msg{block}{B, \var{DAG} \setminus \op{history}[j], i, r}$ to every party. An honest party $j$ receives this message by time $T + \Delta$. Since the message includes all the strong references of $B$, party $j$'s view satisfies $|\var{DAG}(r - 1)| \geq n - f$. As a result, $j$ can create a block for round $r$ once the anchors have sufficient support or a timeout of duration $2\Delta$ triggers. Therefore, $j$ can create its block by time $T + 3\Delta$.
\end{proof}

Lemma~\ref{lemma:follow1} guarantees that committed anchor blocks are in the past of future blocks. The result below generalizes this property to all honest blocks created after the global stabilization time $\var{GST}$.

\begin{lemma}
    \label{lemma:bounds2}
    If honest party $i$ invokes $\op{ab-broadcast}(B,i,r)$ at time $T\geq \var{GST}$, then $B\in\op{past}(B')$ of every honestly $\op{ab-broadcast}(B')$ block $B'$ after time $T+\Delta$
\end{lemma}

\begin{proof}
    Assume that party $i$ invokes $\op{ab-broadcast}(B,i,r)$  at time $T$, then $i$ sends a message of the form $\msg{block}{B, \var{DAG}\setminus\op{history}[j],i,r}$ to every party (L\ref{l:send_history}). Any block $B'$ created by party $j$ after receiving the message $\msg{block}{B, \var{DAG}\setminus\op{history}[j],i,r}$ satisfies $B\in\op{past}(B')$. Lemma~\ref{lemma:bounds1} guarantees that the message $\msg{block}{B, \var{DAG}\setminus\op{history}[j],i,r}$ is not garbage collected, thus proving the statement. 
\end{proof}

\begin{lemma}
    \label{lemma:clock}
    Given an honest party $i$ and a round $r\in\mathbb{N}$, $\exists r'\geq r$ such that $i$ invokes $\op{ab-broadcast}(B,i,r')$ for some block $B$.
\end{lemma}
\begin{proof}

    Let $r$ be a round, and define $r' \geq r$ as the first round such that no honest party has executed it before time $\var{GST}$, and $\op{RR}(r') = i$. We show that party $i$ eventually invokes $\op{ab-broadcast}(B, i, r')$ for some block $B$.

    Consider line~L\ref{l:newround} as executed in round $r'$ by any honest party $j$. Since $i$ is the leader, no honest party concludes round $r$ without a block from $i$ unless its timeout triggers. Consider $j$ to be the first honest party that concluded round $r'-1$, the duration of its timeout is $2\Delta$. This guarantees that party $i$ receives the quorum of round $r-1$ sent by $j$ (at most $\Delta$ delay) before any honest party concludes round $r'$, thus party $i$ invokes ab-broadcast($B,i,r'$)
\end{proof}

\begin{lemma}
    \label{lemma:notimeout}
    At time $T\geq \var{GST}$, if an honest party concludes a round $r$ (L\ref{l:newround}) such that the anchors from rounds $r-2$ and $r-1$ are honest, then $\op{suppAnchor}(r-2)\geq n-f$ and  $\op{suppAnchor}(r-1)\geq n-f$.
\end{lemma}

\begin{proof}
    At time $T \geq \var{GST}$, assume an honest party concludes round $r$ (L\ref{l:newround}) and that the anchors from rounds $r-2$ and $r-1$ are also honest. By construction, if a party creates a block for round $r$ at time $T'$, then every honest party receives this block by time $T' + \Delta$ and subsequently creates its own block, which is received by all honest parties by time $T' + 2\Delta$.

    Given that both anchors from rounds $r-2$ and $r-1$ are honest, it follows that $\op{suppAnchor}(r-2) \geq n - f$ and $\op{suppAnchor}(r-1) \geq n - f$, since there are at least $n - f$ honest parties contributing to support anchors in each round.
\end{proof}

Lemma~\ref{lemma:notimeout} states that after the $\var{GST}$, the timeout is not triggered with honest anchors.

\begin{lemma}
    \label{lemma:latency}
    After the global stabilization time, given a round $r$, the probability that honest party $i$ commits a block in $r$ is at least $\frac{(n-t)^2}{n^2}$. Thus the expected number of rounds until a block is committed is at most $\frac{n^2}{(n-t)^2}$.
\end{lemma}

\begin{proof}
    Given a round $r$, party $i$ commits a block $B \in \var{DAG}(r-2)$ (L\ref{l:anchor}) if and only if the following conditions are satisfied:
    \begin{itemize}
    \item The creator of $B$ has been elected by the round-robin mechanism in round $r-2$, i.e., $\op{creator}(B) = \op{RR}(r-2)$, and the block has sufficient support, $\op{supp}(B) \geq n - f$.
    \item There exists a block $B' \in \var{DAG}(r-1)$ such that $B \in \op{strong}(B')$, $\op{creator}(B') = \op{RR}(r-1)$, and $\op{supp}(B') \geq n - f$.
    \end{itemize}

    Assume that the anchors from rounds $r-2$ and $r-1$ are both honest; this implies the existence of $B\in\var{DAG}(r-2)$ and $B'\in\var{DAG}(r-1)$, as described above. According to Lemma~\ref{lemma:notimeout}, both blocks have enough support: $\op{supp}(B)\geq n-f$ and $\op{supp}(B')\geq n-f$. Since there are $n-t$ honest parties, the anchors from rounds $r-2$ and $r-1$ are both honest with probability $\frac{(n-t)^2}{n^2}$.
   Hence, party $i$ commits a block in round $r$ with probability at least $\frac{(n-t)^2}{n^2}$ and commits every a block every at most $\frac{n^2}{(n-t)^2}$ rounds on average. 

\end{proof}
This lemma shows that, under maximum number of corruptions, honest parties commit a block every at most $9/4$ rounds. Nonetheless, a clever adversary may delay commitment for a short period of time ($t$ rounds), at the cost of allowing the network to commit optimally (every round) outside this period.

\begin{lemma}
\label{lemma:agreement}
    Given two honest parties $i$ and $j$ and a block $B$, then $i$ commits $B$ if and only if $j$ eventually commits $B$.
\end{lemma}

\begin{proof}
    Assume that an honest party $i$ commits a block $B$. By Lemma~\ref{lemma:latency}, any other honest party $j$ will eventually commit a block $B'$ such that $\op{round}(B') \geq \op{round}(B)$. Lemma~\ref{lemma:follow2} applied to blocks $B$ and $B'$ implies that either $B = B'$ or $B \in \op{past}(B')$. Since $\op{round}(B') \geq \op{round}(B)$, it cannot be the case that $B' \in \op{past}(B)$ unless $B = B'$. If $B' = B$, then party $j$ directly commits the same block. If $B \in \op{past}(B')$ and $B \neq B'$, then by construction of the delivery function, party $j$ must have also committed $B$.
    Therefore, we conclude that if an honest party $i$ commits a block $B$, then every honest party $j$ eventually commits $B$.
\end{proof}

\subsection{Main Result}

Previous lemmas are sufficient to show that Black Marlin implements atomic broadcast.

\begin{theorem}
    Black Marlin (Algorithm~\ref{algo:main}) implements atomic broadcast.
\end{theorem}

\begin{proof}
    We proceed property by property:
    \begin{description}
        \item [Validity.] Assume an honest party $i$ invokes $\op{ab-broadcast}(B,i,r)$ after the global stabilization time $\var{GST}$. According to the delivery function (L\ref{l:fun_attempt1}), party $i$ outputs $\op{ab-deliver}(B,i,r)$ the first time it observes a block $B'$ such that $B \in \op{past}(B')$ and $B'$ satisfies line L\ref{l:anchor}.
        Lemma~\ref{lemma:bounds2} guarantees that $B$ is eventually in the past of every valid block. Lemmas~\ref{lemma:clock} and \ref{lemma:latency} ensure that $B'$ is eventually committed, hence party $i$ outputs $\op{ab\text{-}deliver}(B,i,r)$. 

        \item [Agreement.] Suppose an honest party $i$ outputs $\op{ab-deliver}(B,j,r)$. By Algorithm~\ref{algo:main}, this means a block $B^*$ with $B \in \op{past}(B^*)$ satisfies line L\ref{l:anchor}. For any other honest party $j$, Lemmas~\ref{lemma:clock} and \ref{lemma:latency} guarantee that $j$ eventually commits a block $B'$ with $B^*\in\op{past}(B')$, and thus $B\in\op{past}(B')$. Therefore, party $j$ eventually $\op{ab-delivers}(B,j,r)$, ensuring agreement.
              
        \item [Integrity.] The delivery function (L\ref{l:anchor}) ensures that honest party $j$ keeps track of all previously delivered blocks. Consequently, $j$ outputs $\op{ab-deliver}(B,i,r)$ at most once per party $i$ and round $r$. Furthermore, if $i$ is honest, $j$ outputs $\op{ab-deliver}(B,i,r)$ only if $i$ previously invoked $\op{ab\text{-}broadcast}(B,i,r)$ as Byzantine parties cannot impersonate honest parties.
        
        \item [Total order.]  By definition of the delivery function, the order of delivery of non-anchor blocks is determined by the order in which the anchor blocks are committed (L\ref{l:anchor}). Thus, it suffices to show that honest parties commit anchors in the same order. Lemma 12, guarantees that every honest party commits the same set of blocks. The delivery function called with block $B$ guarantees that any anchorin the past of $B$ is committed before $B$,  Lemma 6 guarantees that any committed blocks (possibly by different parties) are in the past of each other. Thus, honest parties commit anchors in the same order.         
    \end{description}
\end{proof}

\subsection{Communication and Time Complexity}
\label{sec:complexities}

Previously, we showed that Algorithm~\ref{algo:main} is both safe and live in a partial synchronous network. In particular, we demonstrated that blocks are eventually delivered, avoiding any concrete analysis of the communication and time complexity of the algorithm. 

\subsubsection{Communication complexity} 
Each round, a party sends at most one message of the form \msg{block}{B, \var{DAG}\setminus\op{history}[j],i,r}. We assume that the round number can be expressed in logarithmic size $\mathcal{O}(\ln (n))$ (note that the round number grows more slowly than the number of blocks). The identifier of the party can also be expressed in $\mathcal{O}(\ln (n))$ number of bits. For the computation of the size of $\var{DAG}\setminus\op{history}[j]$ we assume that the Byzantine parties do not send multiple equivocating blocks, as these blocks include signatures that implicate the party, thus the size of $\var{DAG}\setminus\op{history}[j]\leq \mathcal{O}(n)$. We conclude that the amount of bits sent per honest party and round is $\mathcal{O}(n(|B|+n))$, hence packing $n$ transactions of constant size in $B$ we obtain a communication complexity of $\mathcal{O}(n^2)$ with an amortized complexity of $\mathcal{O}(n)$, both optimal~\cite{DBLP:conf/podc/AbrahamN0X21}.

\subsubsection{Time complexity} Lemma~\ref{lemma:latency} gives an upper bound on the expected number of rounds of communication between anchor blocks are committed. An anchor block is committed on average every $\frac{n^2}{(n-t)^2}$ rounds in the presence of $t$ corruptions: in the worst case, every $9/4$ round of communication, and every round of communication in the good case. Note that in round $r$, an anchor from round $r-2$ is committed. Considering latency to be the number of rounds of communication since an anchor block is broadcast until it is delivered we obtain $4.25$ and $3$ rounds of communication in the average and good case, respectively. It is important to remark that in contrast to most DAG-based protocols, we are measuring in rounds of communication and not instances of reliable broadcast. Thus, Black Marlin not only achieves $\mathcal{O}(1)$ time complexity in expectation, but also the best concrete time complexity concerning DAG-based protocols~\cite{DBLP:journals/iacr/ShresthaKN24,DBLP:conf/fc/SpiegelmanAGL24,DBLP:conf/nsdi/Arun0S0S25,DBLP:conf/sosp/GiridharanSAAC24,DBLP:conf/ndss/BabelCDKKKST25}, as their use of reliable broadcast hinders the protocols.

\section{Benchmarking}
\label{sec:bench}
We evaluate the performance of our implementation with respect to the metrics of throughput and latency for a varying number of parties and network delays. 

\subsection{Experiments}

We conducted all benchmarks locally on a MacBook Pro equipped with an M1 Pro chip, 16 GB of RAM, 10 cores (8 performance and 2 efficiency), and running macOS Sequoia Version 15.0.1. To ensure a fair comparison, we modified the benchmarking framework developed for Bullshark~\cite{DBLP:conf/ccs/SpiegelmanGSK22, bullsharkgit}\footnote{Bullshark has been chosen over more recent frameworks such as Mysticeti~\cite{mysticetigit} due to its closer similarity to Black Marlin and the timing of this project.} to incorporate our implementation of Black Marlin. The biggest difference between Black Marlin and Bullshark is the use of common coins by Bullshark. However, their implementation also uses round-robin\footnote{\url{https://github.com/facebookresearch/narwhal/blob/bullshark/consensus/src/lib.rs} L202.}.
This approach allows a close performance comparison between Black Marlin and Bullshark and also demonstrates that Bullshark’s preprocessing techniques can be adapted to Black Marlin. Note that the actual performance of Bullshark is worse due to the lack of common coins in the implementation.

To emulate realistic network conditions despite running benchmarks on a single machine, we modified both the Black Marlin and Bullshark implementations to introduce configurable network delays. Specifically, messages are delivered based on a Poisson distribution with the expected value $\Delta/2$, where $\Delta$ represents the maximum network delay assumed by the protocol. This design ensures that with high probability, the actual message delay remains below $\Delta$.  For instance, when $\Delta = 400$ ms, the average delivery time for a message is 200 ms.

For each experimental run, we measure \emph{throughput} as the total number of transactions delivered divided by the total duration of the run. The latency of a transaction $\var{tx}$ is defined as the time elapsed from the creation of the block containing $\var{tx}$ until it is \op{ab-delivered}. The \emph{latency} for a run is the average latency of all \op{ab-delivered} transactions in the run. Every party behaves honestly during the benchmarking.

Each data point in Figure~\ref{fig:bench} represents the average throughput or latency over 10 independent runs, each lasting 60 seconds. Error bars indicate two standard deviations. We report results across varying numbers of parties (4, 10, and 13) and network delays ($\Delta = 200$, 400, 600, 800, and 1000 ms). All other parameters remained constant throughout the experiments: one worker per party, a transaction size of 512 bytes, and a transaction injection rate of 50,000 transactions per second.

\subsection{Results}

\begin{figure}[H]

\begin{subfigure}{.5\textwidth}
  \centering
  \includegraphics[width=\linewidth]{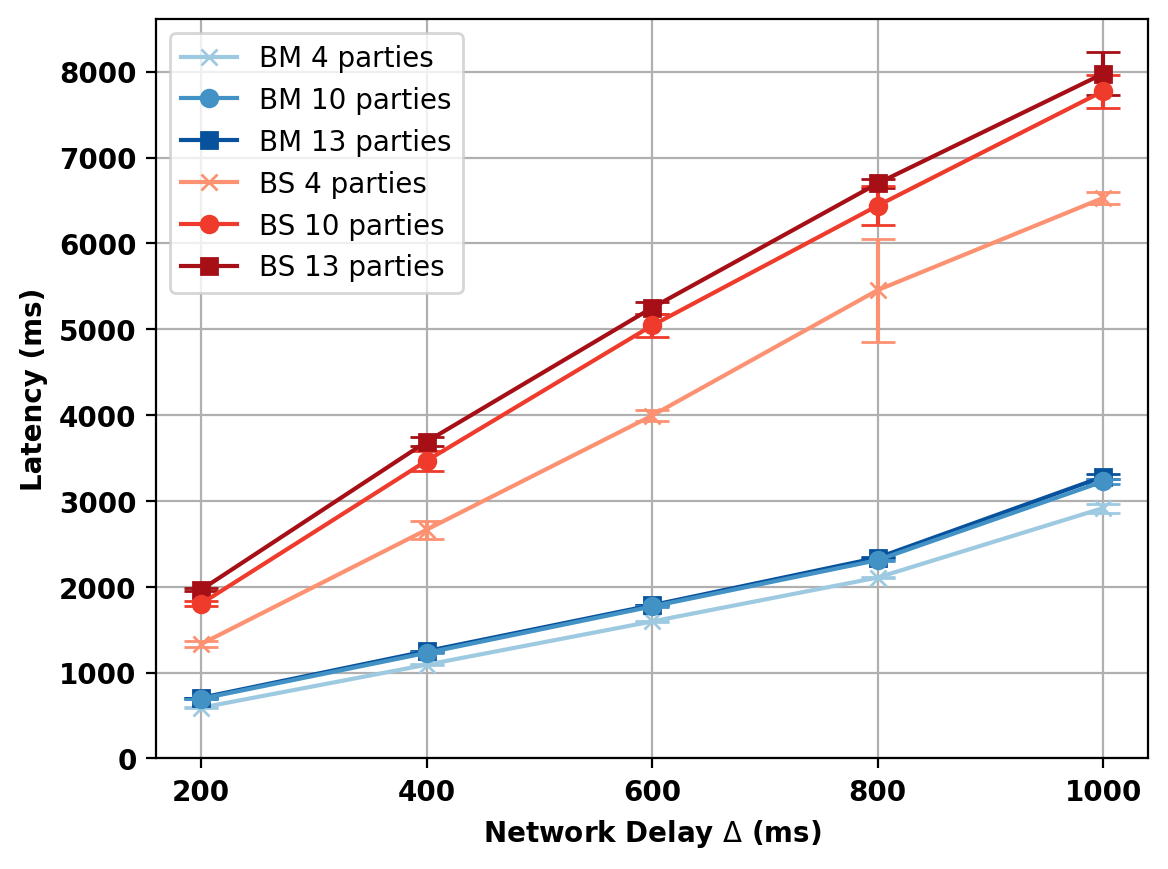}
  \caption{Latency of Black Marlin and Bullshark.}
  \label{fig:sfig1}
\end{subfigure}%
\begin{subfigure}{.5\textwidth}
  \centering
  \includegraphics[width=\linewidth]{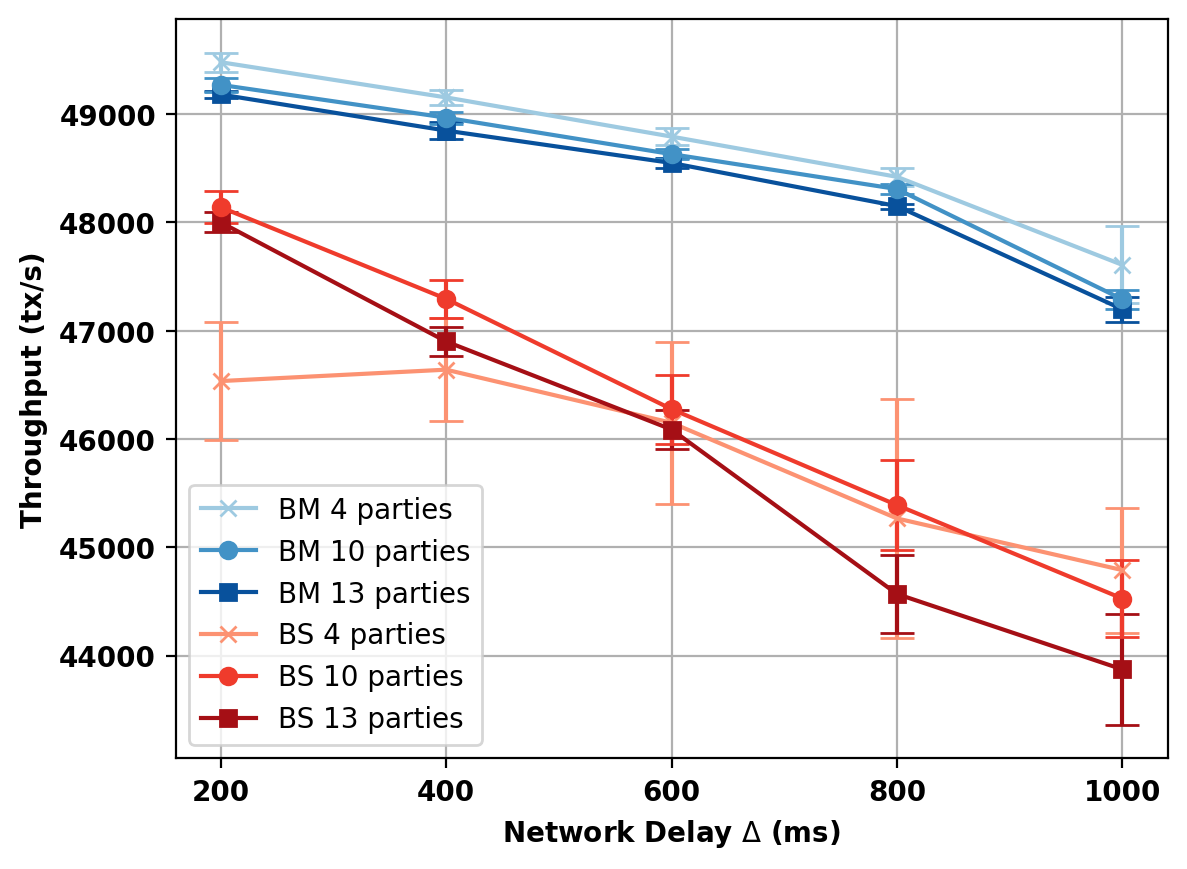}
  \caption{Throughput of Black Marlin and Bullshark.}
  \label{fig:sfig2}
\end{subfigure}\\
\begin{subfigure}{\textwidth}
  \centering
  \includegraphics[width=\linewidth]{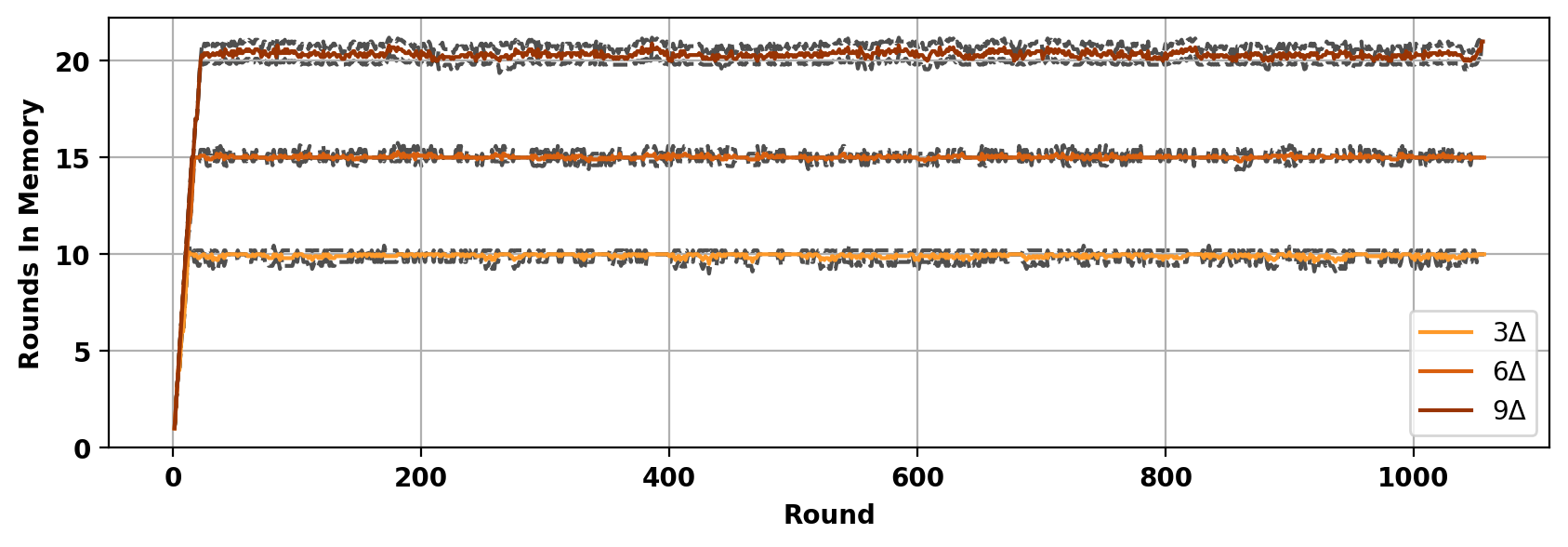}
  \caption{Garbage collection as a function of the estimated network delay.}
  \label{fig:sfig3}
\end{subfigure}
\caption{Black Marlin is shown in blue, Bullshark in red. Darker tones indicate higher numbers of parties: $n = 4$ (cross), $n = 10$ (circle), $n = 13$ (square). Black Marlin outperforms Bullshark in latency and throughput: latency improves by a factor of 2–3; throughput sees a smaller but consistent gain. Bullshark’s block creation strategy explains this modest throughput gap—it produces larger blocks to compensate for delays, up to a size limit. This also accounts for Bullshark’s irregular throughput at $n = 4$ under low delay, where small blocks constrain performance. Plot (c) shows the number of rounds stored in memory by Black Marlin as a function of estimated network delay. The actual delay is $\Delta = 600$ ms, modeled as a Poisson distribution with mean $300$ ms. We vary the garbage collection threshold from $3\Delta$ to $6\Delta$ and $9\Delta$, observing linear growth in retained rounds. As each round contains $\mathcal{O}(n)$ blocks, memory usage remains $\mathcal{O}(n)$.}
\label{fig:bench}
\end{figure}

The results of the experiments described above are summarized in Figure~\ref{fig:bench}, Black Marlin is represented by blue tones and Bullshark by red ones. The two figures of merit that we consider are throughput and latency. 

Black Marlin consistently achieves lower latency than Bullshark across all configurations. The latency improvement ranges from a factor of 2 to 3, depending on the network delay. For example, with 10 parties and a network delay of $\Delta = 200$ ms, Black Marlin achieves a latency of $696 \pm 5$ ms, compared to $1800 \pm 54$ ms for Bullshark. At $\Delta = 1000$ ms, Black Marlin's latency is $3226 \pm 55$ ms, while Bullshark's latency increases to $7773 \pm 360$ ms.
This improvement is expected: Black Marlin eliminates the need for reliable broadcast, reducing latency by approximately a factor of 2. Additionally, its more frequent anchor selection contributes to a constant-factor latency reduction, further widening the latency gap.

Black Marlin also achieves consistently higher throughput than Bullshark across all network delay scenarios. In contrast to the more pronounced latency gains, the throughput improvement is by a constant margin. This behavior can be explained by the way Bullshark constructs blocks: a party running Bullshark accumulates batches of transactions from workers until it is ready to broadcast a new block. As network delay increases, blocks become larger, which helps mitigate throughput loss. Therefore, the $\sim 2\times$ latency improvement observed with Black Marlin does not directly translate into an equivalent gain in throughput—unless the number of workers or the network delay is significantly increased. However, these adjustments were beyond the scope of this study. We were unable to increase the number of workers due to hardware limitations, and we excluded network delays greater than 1 second, as such conditions are considered unrealistic. This block size phenomenon also explains the erratic throughput results observed for Bullshark under configurations with 4 parties and low network delays. In these cases, small block sizes limited Bullshark’s throughput.

Figure~\ref{fig:bench} also demonstrates that Black Marlin’s garbage collection mechanism is robust to overestimations of the actual network delay. Specifically, a constant-factor overestimation leads to only a constant increase in the number of rounds that parties must retain in memory. This implies that even for arbitrarily long executions, Black Marlin only requires $\mathcal{O}(n)$ memory, provided the network delay is overestimated by a constant factor.

Overall, Black Marlin outperforms Bullshark in our implementation, which uses the same codebase and experimental conditions for both protocols. Unfortunately, source code for more recent DAG-based protocols is not publicly available, preventing a direct empirical comparison. However, as we demonstrate in Section~\ref{sec:complexities}, Black Marlin also surpasses these protocols at a theoretical level in terms of communication and latency complexities while allowing implementations with $\mathcal{O}(n)$ memory.

\section*{Acknowledgments}
This work has been mostly funded by the Cryptographic Foundations for Digital Society, CryptoDigi, DFF Research Project 2, Grant ID 10.46540/3103-00077B, and partially funded by the Swiss National Science Foundation(SNSF) under grant agreement Nr. 188443 (Advanced Consensus Protocols) and by a grant from Avalanche, Inc. to the University of Bern.

\bibliographystyle{ieeesort}
\bibliography{references,dblpbibtex}

\end{document}